\documentstyle[amsfonts,epsfig,12pt]{article}
\newcommand{\be}{\begin{equation}}
\newcommand{\ee}{\end{equation}}
\newcommand{\bea}{\begin{eqnarray}}
\newcommand{\eea}{\end{eqnarray}}
\newcommand{\beas}{\begin{eqnarray*}}
\newcommand{\eeas}{\end{eqnarray*}}
\newcommand{\bi}{\begin{itemize}}
\newcommand{\ei}{\end{itemize}}
\newcommand{\bc}{\begin{center}}
\newcommand{\ec}{\end{center}}
\newcommand{\bfl}{\begin{flushleft}}
\newcommand{\efl}{\end{flushleft}}
\newcommand{\bfr}{\begin{flushright}}
\newcommand{\efr}{\end{flushright}}
\newcommand{\f}{\frac}

\def\6{\partial} \def\a{\alpha} \def\b{\beta}
\def\g{\gamma} \def\d{\delta} 
\def\e{\epsilon}
  
  \def\l{\lambda}
\def\m{\mu} \def\n{\nu}  
 \def\s{\sigma} \def\t{\tau}
  
\def\o{\omega} \def\G{\Gamma} 
 \def\L{\Lambda} 
  \def\O{\Omega}

\newcommand{\GG}{{\cal G}}
\newcommand{\HH}{{\cal H}}

\newcommand{\LL}{{\cal L}}

\newcommand{\OO}{{\cal O}}

\begin{document}
\title{First Order Semiclassical Thermal String in the AdS Spacetime}
\author{H. Belich$^{a}$\footnote{belich@cce.ufes.br},
E. L. Gra\c{c}a$^{bc}$\footnote{egraca@ufrrj.br},
M. A. Santos$^{a}$\footnote{masantos@cce.ufes.br}\\
and\\
I. V. Vancea$^{b}$\footnote{ionvancea@ufrrj.br}}
\maketitle

\begin{center}
{\small 
${}^a$
{\em Departamento de F\'{\i}sica e Qu\'{\i}mica, Universidade Federal do Esp\'{\i}rito Santo (UFES)},\\ 
{\em Av. Fernando Ferrari, S/N Goiabeiras, Vit\'{o}ria - ES, 29060-900 - Brasil}\\
${}^b${\em Departamento de F\'{\i}sica, Universidade Federal Rural do Rio de Janeiro (UFRRJ)},\\
{\em BR 465-07-Serop\'{e}dica, RJ, Brasil,}\\
${}^c${\em Centro Brasileiro de Pesquisas F\'{\i}sicas (CBPF)},\\
{\em Rua Dr. Xavier Sigaud 150, 22290-180 Rio de Janeiro, RJ, Brasil,}\\
}
\end{center}

\begin{abstract}
We formulate the finite temperature theory for the free thermal excitations of the bosonic string in the anti-de Sitter (AdS) spacetime in the Thermo Field Dynamics (TFD) approach. The spacetime metric is treated exactly while the string and the thermal reservoir are semiclassically quantized at the first order perturbation theory with respect to the dimensionless parameter $\epsilon = \a ' H^{-2}$. In the conformal $D=2+1$ black-hole AdS background the quantization is exact. The method can be extended to the arbitrary AdS spacetime only in the first order perturbation. This approximation is taken in the center of mass reference frame and it is justified by the fact that at the first order the string dynamics is determined only by the interaction between the {\em free} string oscillation modes and the {\em exact} background. The first order thermal string is obtained by thermalization of the $T = 0$ system carried on by the TFD Bogoliubov operator. We determine the free thermal string states and compute the local entropy and free energy in the center of mass reference frame.
\end{abstract}

\newpage

\section{Introduction}

The bosonic string in the anti-de Sitter (AdS) spacetime represents the first example of an exactly quantized string theory on a curved spacetime manifold. As such, its classical and quantum dynamics have been extensively studied since the pioneer works of \cite{brfw,dpl}.  Some of the problems of the string theory in the AdS spacetime which have attracted more attention during the years concern the understanding of the string spectrum, the unitarity of the theory and, more recently, the implications of the conjectured relation between the superstring and the supergravity in the AdS spacetime \cite{jm}.   

The differences between the string dynamics in the AdS and the Minkowski spacetimes are due to the fact that, in general, the $D$-dimensional AdS spacetime is not a solution of the full $\b$-function equations for the string $\sigma$-model. Therefore, there are large classes of conformal and non-conformal AdS backgrounds in which the physical properties of the quantum string are difficult to study. The conformally invariant backgrounds are necessary to define consistent quantum string theories. However, many backgrounds interesting from the physical point of view (among which some are derived from the AdS spacetime) do not satisfy this requirement.  

Some time ago, a method for analysing the bosonic string dynamics in general spacetime was proposed in \cite{ns1,ns2,nsa}. There it was shown that by choosing appropriate boundary conditions for the bosonic string in an arbitrary spacetime, the reparametrization invariance of the world-sheet theory can be recasted into coordinate transformations between different frames in spacetime. The world-sheet holomorphic mappings depend on the fundamental string length which is a new parameter and determines, e. g. the periodicity of the closed string in the non-inertial frames. Also, a local light-cone gauge can be chosen in any non-inertial frame. In this gauge one can locally split the string degrees of freedom into longitudinal, i. e. along the trajectory of the string center of mass, and transversal, and show that the longitudinal degrees of freedom are functions of the transversal ones. These facts led the authors of \cite{ns1,ns2} to developing an approximation scheme for the canonical quantization of string theory in the conformally invariant backgrounds called {\em semiclassical quantization} in which the metric is kept fixed while the perturbation is performed around the trajectory of the center of mass \cite{ns2,ns3,ns4,ns6}. In the same papers the semiclassical quantization method was extended at the first order to the non-conformal $D$-dimensional AdS background. The main idea behind this generalization was the observation that the conformally invariant $SL(2,R)$-WZWN background can be mapped into the AdS spacetime with a non-vanishing parallelizing torsion by a point identification and a reparametrization. (Previously, it had been shown in \cite{btz} that the $(2+1)$-dimensional black-hole AdS background is an exact solution of the string theory.) Since the first order string perturbations and the conformal generators from the theory defined in the conformally invariant $(2+1)$ black-hole AdS spacetime do not depend on the black-hole mass and this background is asymptotically AdS, the results were naturally generalized to the non-conformal $D$-dimensional AdS spacetime. 

One of the most attractive features of the semiclassical quantization is that it allows to study the {\em free transverse} quantum string excitations interacting with the {\em exact background metric}. This is achieved by working at the first order in the power expansion of the string coordinates with respect to a dimensionless parameter which in the $D=2+1$ black hole AdS background is $\epsilon = \a ' l^{-2}$ and $l$ is the typical length scale.
Starting from the second order, the expansion coefficients display the interaction among the oscillators and with the background. However, the expansion is still controlled by ordinary differential equations \cite{ns3}. The quantization is performed in a reference frame defined by the zeroth order term in the power expansion which is required to be a solution of the geodesic equations, i. e. the trajectory of the string center of mass. The transverse string oscillating modes can be quantized by canonical methods if one chooses a covariantly constant basis for the polarization vectors. The longitudinal excitations differ from one reference frame to other. However, the longitudinal degrees of freedom can be expressed in terms of transversal degrees of freedom by using the vanishing of the energy-momentum tensor equation \cite{ns1}. This method describes the low excitations of the bosonic string as compared with the energy scale of the background metric.  In the black-hole background it corresponds to black-hole masses larger than the string energy, i. e. to a strong gravitational field \cite{ns2}.

The exact anomaly free quantization is guaranteed only in the $D=2+1$ black hole AdS spacetime. However, the solutions of the classical equations of motion can be expanded around the geodesic of the center of mass in the arbitrary dimension AdS spacetime. The parameter of the expansion in this case is $\epsilon = \a ' H^{2}$. Here, $\a '= (2\pi T_s )^{-1}$, $T_s$ is the string tension and $H = l^{-1}$ is the Hubble constant related to the cosmological constant $\L$ by $\L = - (D-1)(D-2)H^{2}/2$. Since the first order string excitations and the conformal generators in $D=2+1$ do not depend on the black hole parameters, one can reconstruct them from the corresponding first order quantities in the $D$-dimensional AdS spacetime. In this case, the transverse string excitations are spanned by $(D-1)$ physical polarizations which introduce an $SO(D-1)$ gauge group. Thus, the interpretation of the $D=2+1$ quantum theory can be maintained in the $D$-dimensional AdS spacetime as long as we limit the analysis to the first order terms only \cite{ns6}, i. e. the spacetime typical radius is much larger than the Planck scale. (Hence the explanation of the adjective of semiclassical.) If higher order terms are taken into account, the quantization method is no longer valid since these terms contain the interaction among string modes and between the string and the metric. Thus, the physical string states are no longer fixed by the generators of the conformal symmetry alone which is another way of seeing the breaking of the conformal symmetry. The semiclassical quantization has proved its usefulness in investigating problems such as the string dynamics in black hole, cosmological, gravitational wave and supergravity backgrounds (see \cite{ns4} and the references therein.) 

In this paper we are going to construct the {\em thermalized free excitations} of the bosonic string in the black-hole AdS spacetime \cite{bhtz} within the TFD framework \cite{tu,ubook}. The bosonic string partition function in this background  was discussed in the past by using the Matsubara formalism (see e. g. \cite{ns6}). However, our goal is different in that by applying the TFD method we provide the thermal free string states in the unperturbed background. Since the free string excitations are available in the first order perturbation of the semiclassical quantization, in our approach we are going to consider this approximation. However, note that our results are valid in the full quantum theory in $D=2+1$ black-hole AdS background as well as in $SU(2,R)$ background. In this case, the TFD method shows the breaking of the conformal symmetry at finite temperature. Our first order construction in arbitrary $D$-dimensional AdS background is justified by the extension of the semiclassical quantization method in the first order to the non-conformal AdS background as discussed above. However, the results are formally the same in the two cases. The breaking of the conformal invariance in the first order semiclassical quantization is irrelevant in $D>2$ since the full theory is not conformally invariant. Also, we are going to calculate the entropy and the free energy of the thermal string in the string center of mass reference frame. 

The TFD was employed previously in the study of the thermal string field in the Minkowski spacetime in \cite{yl1}-\cite{yl6} and the first quantized string and superstring in \cite{fns1}-\cite{f1}. 
More recently, the thermal bosonic open string states were discussed in \cite{ivv7}. In \cite{ng1,ng2} the string entropy from the TFD in the $pp$-wave background was calculated. An important result was obtained in \cite{ng3} where it was shown the equivalence between the TFD and the Matsubara formalisms, respectively, in the string theory (see also \cite{ng4,ng5}.) In \cite{ivv1,ivv2,ivv3}, a new method for obtaining thermal boundary states from bosonic $D$-brane states was proposed (see for reviews and related problems \cite{ivv4,ivv5,ivv6}.) The method was generalized to the supersymmetric Green-Schwarz superstring for which thermal boundary states were obtained from the Bogomolny-Prasad-Sommerfeld D-branes \cite{ivv8,ivv9}. The present paper represents a generalization of the previous works and opens the possibility to use the TFD to study locally the microscopic formulation of the thermal $D$-branes in the AdS and black-hole AdS backgrounds in an perturbative approach. (For interesting recent discussions of the perturbative string in the AdS spacetime and the calculation of the string partition function using the random walks method in static backgrounds see \cite{fg,kl}.) 

The paper is organized as follows. In the next section we review the semiclassical quantization of the bosonic string in the black-hole AdS background. We thermalize the string in the first order of the power expansion around the geodesic in Section III. Here, we derive the general form of the thermal string states. In Section IV we compute the entropy and the free energy of the thermal bosonic string. The last section is devoted to discussions.

\section{Semiclassical Quantization of the Bosonic String in the Black-Hole AdS Background}

In this section we are going to review the semiclassical quantization of the bosonic string in the black-hole AdS background following \cite{ns3}. The results are exact for the full quantum theory in the conformal $D=2+1$ black-hole AdS background. However, as shown in \cite{ns6}, they can be extended to the arbitrary dimensional AdS background in the first order of the perturbation series to be given below. Since our approach is in this approximation in which the results are formally the same regardless the dimension of the AdS, our notation will be for the arbitrary AdS spacetime. For details of the extension from $D=2+1$ to arbitrary $D$ we reefer the reader to the paper \cite{ns6}.  

The action of the bosonic string in the $D$-dimensional AdS spacetime is given by the following functional
\be
S = \frac{1}{2 \pi \a '} \int d^2 \s \sqrt{h} h^{\a \b} g_{ab}(x) \6_{\a}x^{a} \6_{\b}x^{b}, 
\label{action}
\ee
where $\s = (\s^0 , \s^1 )$ are the world-sheet coordinates, $h_{\a \b}$ is the world-sheet metric, $g_{ab}(x)$ is the spacetime metric and $x^{a}(\s^0 ,\s^1 )$ are the string coordinates in the spacetime. Here, $\a , \b = 1,2$ are the world-sheet indices and $a, b, = 1, 2, \ldots , D$ are the spacetime indices. The energy-momentum tensor is obtained by deriving the action (\ref{action}) with respect to the world-sheet metric
\be
T_{\a \b} \equiv \frac{2}{\sqrt{h}}\frac{\delta S}{\delta h^{\a \b}} = g_{ab}(x)\left( \6_{\a}x^{a} \6_{\b}x^{b} - \frac{1}{2} h_{\a \b} \6_{\g}x^{a} \6^{\g}x^{b} \right).
\label{energymomentum}
\ee
By the equations of motion of $h^{\a \b}$ the energy-momentum tensor vanishes $T_{\a \b} = 0$. One can use the reparametrization invariance of the action to fix the world-sheet conformal symmetry
\be
h_{\a \b}(\s ) = e^{\L(\s)}\eta_{\a \b}. 
\label{conformalgauge}
\ee 
In the conformal gauge the classical string dynamics is determined by the equations of motion and the constraints arising from the vanishing of the energy-momentum tensor
\bea
\ddot{x}^{a} - x''^{a} + \G^{a}_{bc}(x)\left( \dot{x}^b\dot{x}^c - x'^{b}x'^{c} \right)&=&0,
\label{eqmotion}\\
g_{ab}(x)\dot{x}^a x'^{b} = g_{ab}(x)\left( \dot{x}^a\dot{x}^b + x'^{a}x'^{b} \right)&=&0,
\label{constraints}
\eea 
where $\s^{\a} = (\t , \s )$. 

The semiclassical quantization method was developed for studying the quantum string excitation in the exact classical background \cite{ns1,ns2,ns3}. Its starting point is expanding the string fields $x^{a}(\t , \s )$ around the exact solution of the equation of motion of its center of mass (the geodesic) denoted by $\eta^{a}_{0}(\t )$
\be
x^{a}(\t , \s ) = \sum^{\infty}_{n = 0}\e^n \eta^{a}_{n}(t, \s),
\label{fieldexpansion}
\ee
with the initial condition $\eta^{a}_{0}(\t , \s ) = \eta^{a}_{0}(\t )$ satisfying the following equations of motion and constraint
\bea
\ddot{\eta}^{a}_{0} + \G^{a}_{bc}(\eta_{0})\dot{\eta}^{b}_{0}\dot{\eta}^{c}_{0} &=& 0,
\label{eqmotioncm}\\
g_{ab}(\eta_0)\dot{\eta}^a_{0} \dot{\eta}^{b}_0 &=& -m^2 {\a '}^2.
\label{constraintcm}
\eea 
The relation (\ref{constraintcm}) defines the string mass operator which spectrum was discussed in \cite{ns1}. In the $D$-dimensional AdS spacetime there are $D-1$ physical polarizations of string perturbation around the geodesic $\eta^{a}_{0}$. Consequently, $D-1$ transverse normal vectors $n^{a}_{\m}$, $\m = 1, 2, \ldots , D$ can be introduced
\bea
g_{ab}(\eta_0 ) n^{a}_{\m}\dot{\eta}^{b}_{0} &=& 0,
\label{normalitycond}\\
g_{ab}(\eta_0 ) n^{a}_{\m} n^{b}_{\n} &=& \d_{\m \n}.
\label{ortonormalitycond}
\eea
The choice of the set $ \{ n^{a}_{\m} \}$ is not unique since there is a local gauge group $SO(D-1)$ corresponding to the rotations of $ \{ n^{a}_{\m} \}$. This gauge symmetry is fixed by requiring that the normal vectors be covariantly constant
\be
\dot{\eta}^{a}_{0} \nabla_a n^{b}_{\m} = 0.
\label{covconst}
\ee
In this gauge most of relations among the normal vectors take a simpler form. In particular, the vectors from $\{ n^{a}_{\m} \}$ satisfy the following completeness relation
\be
g^{ab} = -\frac{1}{m^2}\dot{\eta}^{a}_{0}\dot{\eta}^{b}_{0} + n^{a}_{\m}n^{b}_{\n}\d^{\m\n}.
\label{completenessrel}
\ee
The first order co-moving string perturbations from the expansion (\ref{fieldexpansion}) can be written in the following form
\be
\eta^{a}_{1}(\t , \s ) = \d x^{\m}(\t ,\s )n^{a}_{\m}.
\label{firstordperturb}
\ee 
In the non-rotating black-hole AdS spacetime, the solution of the equations of motion (\ref{eqmotion}) in the first order in $\e $ has the following Fourier expansion \cite{ns1}
\be
\delta x^{\m} (\t , \s ) = \sum_{n \neq 0 } 
\sqrt{\frac{2|n|\O_{n}}{\a ' }}
\left[ \a^{\m}_{n} e^{-in(\O_n \t - \s)} + \b^{\m }_{n}e^{-in(\O_n \t + \s)} \right]
+ \sqrt{\frac{l}{2m}} \left[ \a^{\m}_{0} e^{-i\frac{m\a '}{l}\t} + 
\b^{\m }_{0}e^{+i\frac{m\a '}{l}\t} \right].
\label{Fourierexp}
\ee
Here, the frequencies of the string oscillators in $ \hbar = 1$ units are $\o_0 = m\a ' / l$ and $\o_n = \o_{-n} = |n| \O_n$ for $ n = \pm 1, \pm 2, \ldots$. Also, we are using the following notation
\be
\O_n = \sqrt{1 + \frac{m^2 {\a '}^2}{n^2 l^2}}.
\label{Omega}
\ee
It can be shown that the first order perturbations from the relation (\ref{Fourierexp})
satisfy the equation of motion derived from the action given in the relation (\ref{action}) truncated at second order \cite{ns1}
\be
S_{2} = - \frac{1}{2\pi\a '} \int d\s d\t \sum_{\m =1}^{D-1} \left( \eta^{\a\b}
\6_{\a} \delta x^{\m} \6_{\b} \delta x^{\m} + \frac{m^2 {\a '}^2}{l^2} \d x^{\m} \d x_{\m}
\right). 
\label{truncatedaction}
\ee
The next step is to impose the constraints from the relation (\ref{constraints}). This can be done by solving the equations of motion (\ref{eqmotion}) for the first and second order perturbations \cite{ns1}. In the world-sheet light-cone gauge $\s^{\pm}=(\t \pm \s)$ the lower order components of the energy-momentum tensor have the following form
\be
T_{\pm\pm} = g_{ab}\6_{\pm}x^a\6_{\pm}x^b=0.
\label{lowerenergmom}
\ee
In the $D=2+1$, the contribution from $T_{\pm \mp} =0$ by the conformal invariance. Since the 
components $T_{\pm\pm}$ are conserved, they can be written as
\bea
T_{--} = \frac{1}{2 \pi }\sum\limits_{n} L^{-}_{n} e^{-in(\s - \t)},
\label{T--}\\
T_{++} = \frac{1}{2 \pi }\sum\limits_{n} L^{+}_{n} e^{-in(\s + \t)}.
\label{T++}
\eea  
The coefficients $L^{-}_{n}$ and $L^{+}_{n}$ are the conformal generators. They can be expressed in terms of the Fourier coefficients of the first order perturbations given in the relation (\ref{Fourierexp}). Thus, the classical constraints have the form 
\be
L^{-}_{n}=L^{+}_{n}=0.
\label{classconstrL}
\ee
In $D=2+1$ AdS spacetime the string mass operator can formally be obtained from $L^{-}_{0}$ and $L^{+}_{0}$ which have the following form
\bea
L^{-}_{0} & = & \pi \a ' \sum\limits_{n > 0}
\left[ \frac{(\o_n - n)^2}{2n \O_n } \b^{\dagger}_{n} \cdot \b^{\dagger}_{n} +
\frac{(\o_n + n)^2}{2n \O_n } \a^{\dagger}_{n} \cdot \a^{\dagger}_{n} \right] +
\frac{\pi m {\a '}^2}{2l} \a^{\dagger}_{0} \cdot \a_{0} - 
\frac{\pi m^2 {\a '}^2}{2},
\label{L-}\\
L^{+}_{0} & = & \pi \a ' \sum\limits_{n > 0}
\left[ \frac{(\o_n + n)^2}{2n \O_n } \b^{\dagger}_{n} \cdot \b^{\dagger}_{n} +
\frac{(\o_n - n)^2}{2n \O_n } \a^{\dagger}_{n} \cdot \a^{\dagger}_{n} \right] +
\frac{\pi m {\a '}^2}{2l} \a^{\dagger}_{0} \cdot \a_{0} - 
\frac{\pi m^2 {\a '}^2}{2},
\label{L+}
\eea
where $\cdot$ stands for the sum over $\m = 1, 2$. Upon quantization, the Fourier coefficients are turned into operators that obey the standard commutation relations
\be
[\a^{\m}_{m}, \a^{\dagger\n}_{n} ] = [\b^{\m}_{m}, \b^{\dagger\n}_{n} ] = \d^{\m\n}\d_{mn}
~~,~~
[\a^{\m}_{m}, \b^{\n}_{n} ] = 0
~~,~~
[\a^{\m}_{0}, \a^{\dagger\n}_{0} ]  =  \d^{\m\n},
\label{osccommrel}
\ee
where 
$\a^{\m}_{-n} = \a^{\dagger\m}_{n}$, $\a^{\m}_{-n} = \a^{\dagger\m}_{n}$, 
and $\b^{\m}_{0} = \a^{\dagger\m}_{0}$. As a matter of notation, in what follows the quantities with an over bar will denote the $\b$ oscillators, while the ones without an over bar will denote the $\a$ oscillators. The constraints arising from zero modes generators are
\be
\left( L^{-}_0 - 2 \pi \a ' \right) \left| \Psi_{phys} \right\rangle = 0
~~,~~
\left( L^{+}_0 - 2 \pi \a ' \right) \left| \Psi_{phys} \right\rangle = 0,
\label{constr}
\ee
The generators of the world-sheet translations $\s \rightarrow \s + \xi$ and $\t \rightarrow \t + \zeta$ are $P = L^{+}_{0} - L^{-}_{0}$ and $H = L^{+}_{0} + L^{-}_{0}$, respectively. By introducing the number operators 
\be
N_n  =  \frac{n}{2}\sum\limits^{D-1}_{\m = 1}\a^{\dagger \m}_{n} \a^{\m}_{n}
~~,~~
\overline{N}_n  =  \frac{n}{2}\sum\limits^{D-1}_{\m = 1}\b^{\dagger \m}_{n} \b^{\m}_{n},
\label{numberNbarN}
\ee
for $\a_{n}$ and $\b_{n}$ oscillators, respectively, we obtain the following relations
\bea
H & = & 2\pi\a ' \sum\limits_{n \geq 1}
\left( \frac{\O^{2}_{n} + 1}{\O_n} \right) \left( N_n + \overline{N}_{n}\right)
+\frac{\pi m {\a '}^2}{l} \a^{\dagger}_{0} \cdot \a_{0} - 
\pi m^2{\a '}^2,
\label{hamiltonian}\\
P & = & 4\pi\a ' \sum\limits_{n \geq 1}\left( N_n - \overline{N}_{n}\right).
\label{momentum}
\eea
The momentum constraint takes the form of the level matching condition for the closed bosonic string
\be
4\pi\a ' \sum\limits_{n \geq 1}\left( N_n - \overline{N}_{n}\right)
\left| \Psi_{phys} \right \rangle = 0.
\label{levelmatching}
\ee
The constraints (\ref{constr}) and (\ref{levelmatching}) must be imposed on the Hilbert space of the bosonic string in order to project the dynamics in to the physical subspace.

In the arbitrary AdS spacetime the unitarity is achieved by imposing spin-level restrictions on the allowed representations of the Virasoro algebra (see e. g. \cite{egp} and the references therein), beside the Virasoro constraints, since the later in general do not eliminate completely the negative norm state \cite{brfw}. However, since the formulas obtained above for the $D=2+1$ black-hole AdS do not depend on the black-hole mass, and this background is asymptotically AdS, following \cite{ns6} we generalize the above relations to the arbitrary AdS in the first order of perturbation from (\ref{fieldexpansion}) and take $\m, \n =1, 2, \ldots, D-1$ in the above formulas. It is important to note, however, that $L^{+}_n$ and $L^{-}_{n}$ do not generate an exact symmetry in the arbitrary AdS spacetime. Nevertheless, the corresponding Hamiltonian and level matching condition operators are obtained from the generalization of the $D=2+1$ case. Therefore, by an abuse of notation, we will use $P$ for level matching condition in what follows.   

Note that in the AdS spacetime the string excitations oscillate in time. However, the possible instabilities do not develop due to the negativeness of the local gravity. The string states can be organized as eigenstates of the mass operator which is obtained from the constraints (\ref{constr}). For more details on this construction we refer the reader to \cite{ns1,ns2,ns6}.

\section{Thermal String in the TFD Approach}

In order to obtain the first order thermal bosonic string we are going to apply the TFD formalism \cite{tu,ubook} to the semiclassicaly quantized string from the previous section. In the first subsection we are going to discuss the TFD ansatz for the bosonic string and calculate its thermal partition function $Z(\b_T)$ and the thermal vacuum $| 0(\b_T ) \rangle\rangle $, where $\b_T = (k_B T)^{-1}$ and $k_B$ is the Boltzmann constant. In the calculation of $Z(\b_T)$ there are formal differences between working in the total Hilbert space $\HH$ and the physical subspace $\HH_{phys} \in \HH$, respectively. From the form of the partition function in the physical subspace we conclude that the Bogoliubov operator is of the known form and the string thermalization can be performed straightforwardly. The general form of the thermal string states is given.

\subsection{Thermal String Vacuum}

The thermalization is the physical process in which a system is put in contact with a thermal reservoir to heat it up from $T = 0$ to some $ T > 0$. In the TFD formalism, this process is described as follows \cite{tu,ubook}. Each independent quantum oscillation mode of the initial system interacts with an identical degree of freedom of the reservoir. The specific interaction is expressed by the Bogoliubov operator which mixes the pair of oscillators. The result are two new degrees of freedom which are temperature dependent. Thus, the heating of the original system is the result of the interaction of all oscillators one by one with identical degrees of freedom of the reservoir and the dynamics of the finite temperature system is described in terms of new pairs of oscillators that mix zero temperature system and reservoir oscillating modes, respectively. Isolating the system oscillators and the reservoir oscillators is called {\em doubling the system}. 

The TFD ansatz states that the statistical average of any observable $O$ should be expressed as an expectation value in a state $|0(\b_T)\rangle\rangle$  that characterizes the system at equilibrium at finite temperature called the {\em thermal vacuum}. This statement is expressed by the following relation  
\be
\langle O \rangle = Z^{-1}(\b_T)\mbox{Tr}\left[ e^{-\b_T H} O \right] \equiv
\langle\langle 0(\b_T)|O|0(\b_T)\rangle\rangle,
\label{TFDansatz}
\ee
where $H$ is the Hamiltonian of the system. As discussed in \cite{ng3,ng4,ivv8}, the above ansatz should be modified when applying the TFD to string theory to the following relation
\be
\langle O \rangle = Z^{-1}(\b_T)\mbox{Tr}\left[ \d(P=0) e^{-\b_T H} O \right] \equiv
\langle\langle 0(\b_T)|O|0(\b_T)\rangle\rangle,
\label{TFDansatzstring}
\ee
which takes into account the world-sheet reparametrization invariance. The relation (\ref{TFDansatzstring}) holds after the string symmetries have been fixed. It implies that only the physical states contribute to the trace. In a more general situation, the full set of constraints should be implemented into the trace in order to guarantee that it is calculated in the physical subspace $\HH_{phys}$ and not in the full string Hilbert space $\HH$.

Let us apply these ideas to the bosonic string in the $D=2+1$ black-hole AdS background from the previous section. Denote all the physical quantities and the degrees of freedom of the thermal reservoir that correspond to the string oscillating modes by a tilde. Then the total system at $T=0$ has the full Hilbert space $\hat{\HH} = \HH \otimes \tilde{\HH}$. According to the TFD, the thermal vacuum state $|0(\b_T)\rangle\rangle $ should have the following form
\be
|0(\b_T )\rangle\rangle = \sum\limits_{w}\sum\limits_{\overline{w}}f_{w,\overline{w}}(\b_T)|w\rangle|\overline{w}\rangle ,
\label{thermvacexp}
\ee  
where $w$ and $\overline{w}$ are multi-indices corresponding to the $\a$ and $\b$ modes, respectively. We introduce the following notations for the eigenvalues of the number operators
\be
N_n = n \sum\limits_{\m = 1}^{D-1}k^{\m}_{n}~~,~~
\overline{N}_n = n \sum\limits_{\m = 1}^{D-1}\overline{k}^{\m}_{n},
\label{Neigenvalues1}
\ee
where $k^{\m}_{n}, \overline{k}^{\m}_{n}$ are eigenvalues of the number operators $N^{\m}_{n} = \a^{\dagger\m}_n \a^{\m}_n$ and $\overline{N}^{\m}_{n} = \b^{\dagger\m}_n \b^{\m}_n$, respectively, for any $\m = 1, 2, \ldots , D-1$ and $n = 1, 2, \ldots$ i. e. they satisfy the following relations
\be
N^{\m}_{n}| \cdots k^{\m}_{n} \cdots \rangle = n k^{\m}_{n} | \cdots k^{\m}_{n} \cdots \rangle ,
~~,~~
\overline{N}^{\m}_{n}| \cdots \overline{k}^{\m}_{n} \cdots \rangle = n \overline{k}^{\m}_{n} | \cdots \overline{k}^{\m}_{n} \cdots \rangle .
\label{Neigenvalues2}
\ee
Then the following orthogonality relation holds
\bea 
f^{*}_{w',\overline{w}'}(\b_T)f_{w,\overline{w}}(\b_T) & = &
Z^{-1}(\b_T)\d(w',w)\d(\overline{w}',\overline{w})
\frac{\exp\left(\b_T \pi m^2 {\a '}^2 \right)}
{\left[ 1 - \exp \left( -\frac{\b_T \pi m {\a '}^2}{l} \right) \right]^{D-1}} \times
\nonumber\\
& & \int\limits_{-1/2}^{+1/2} ds \exp\left[ 2\pi \a ' \sum\limits_{n}
\left( \overline{\l}_n(\b_T ,s)\overline{N}_n + {\l}_n(\b_T ,s) N_n \right)\right],
\label{orthogf}
\eea
where $\d(w',w)$ and $\d(\overline{w}',\overline{w})$ are short hand notations for the product of delta functions for each pair of indices in the corresponding multi-index and  
\be
{\l}_n(\b_T ,s) = - \b_T \o_n - \f{is}{\a '}~~,~~
\overline{\l}_n(\b_T ,s) = -\b_T \o_n + \f{is}{\a '}.
\label{lambdas}
\ee
The constraint has been written using the following analytic representation of the delta function
\be
\delta(P=0)\equiv\delta(\overline{N} - N) = \int\limits^{+1/2}_{-1/2} ds \,
e^{2\pi i s\left(\overline{N} - N\right)}.
\label{deltaanalytic}
\ee
The orthogonality relation (\ref{orthogf}) shows that, as in the case of the Minkowski spacetime, the coefficients in the expansion of the thermal vacuum are vectors from a Hilbert space identical with the string Hilbert space, i. e. the Hilbert space $\tilde{\HH}$ of the reservoir degrees of freedom. Also, the relation (\ref{orthogf}) shows that in the expansion of $|0(\b_T )\rangle\rangle$ in the full Hilbert space $\hat{\HH} = \HH \otimes \tilde{\HH}$, these vectors are tensored with Columbeau functionals \cite{c}, i. e. the square root of the delta function. This suggests that the thermal vacuum is actually a state from the total physical space $\hat{\HH}_{phys} \in \hat{\HH}$ rather than the full space. Indeed, there is no delta function factor if the trace from (\ref{TFDansatzstring}) is taken over  $\hat{\HH}_{phys}$ instead of $\hat{\HH}$ and, consequently, there is no dependence of the thermal vacuum on the constraints. For simplicity, we will work in what follows in the physical space. Then the relation (\ref{thermvacexp}) takes the following form
\bea
|0(\b_T )\rangle\rangle & = &
Z^{-\frac{1}{2}}(\b_T)
\d(w',w)\d(\overline{w}',\overline{w})
        \frac{\exp 
                  \left( 
                        \frac{\b_T \pi m^2 {\a '}^2}
                             {2} 
                   \right)
             }
             {\left[ 1 - \exp \left( 
                                    -\frac{\b_T \pi m {\a '}^2}
                                          {l} 
                               \right)
              \right]^{\frac{D-1}{2}}
             }\times \nonumber\\
& & \sum\limits_{w}\sum\limits_{\overline{w}}
\exp\left[ -\b_T \pi \a ' \sum\limits_{n=1}^{\infty} \o_n \left( \overline{N}_n + N_n \right)\right]
|w,\overline{w}\rangle
\widetilde{|w,\overline{w}\rangle}.
\label{thermvacphs}
\eea
Here, the multi-index states are $|w,\overline{w}\rangle \in {\HH}_{phys}$
and $\widetilde{|w,\overline{w}\rangle} \in {\tilde \HH}_{phys}$, respectively. The partition function can be obtained by requiring that the norm of the thermal vacuum be equal to one and taking the trace of the identity operator in the physical subspace
\be
Z(\b_T) = 
\frac{\exp \left( \b_T \pi m^2 {\a '}^2 \right)}
{\left[ 1 - \exp \left(-
                       \frac{ \b_T \pi m {\a '}^2}
                            {l} 
                 \right)
              \right]^{D-1}
             }
\prod\limits_{n=1}^{\infty}
\left( 1- e^{-\b_T \pi \a ' n \o_n} \right)^{2(1-D)}.
\label{partfuncphys}
\ee             
Here, the factor of 2 in the exponential comes from the equal contributions of both $\a$ and $\b$ oscillators, respectively.

In the arbitrary AdS spacetime, there is no need to impose the momentum constraint in the calculation of the trace. However, since the level matching condition still holds at first order in the expansion (\ref{fieldexpansion}), we impose it in the general case, too. Thus, the above considerations are valid for arbitrary AdS in the approximation considered.
The relations (\ref{thermvacphs}) and (\ref{partfuncphys}) show that the decomposition of the thermal vacuum in terms of physical states is similar to that of a free quantum field in the Minkowski spacetime. However, there are two important differences. The first one is the contributions of zero mode and square mass to the exponential. The second one is deeper and concerns the validity of the above state as a thermal string vacuum only locally, in the free falling reference frame of the center of mass, i. e. along the geodesics of the AdS spacetime.

\subsection{Thermal String States}

The form of the thermal vacuum and the partition function obtained in the previous subsection show that the TFD program can be carried to the semiclassical string bosonic string in the AdS background. The thermal string states belong to the Hilbert space $\HH(\b_T )$ and are generated by thermal creation operators acting on the thermal vacuum. The mapping from the theory at $T = 0$ to the theory at $T \neq 0 $ is generated by the temperature dependent Bogoliubov operator corresponding to all oscillators of the total system
\be
\GG = G_0 + G + \overline{G},
\label{Bogoliubovop}
\ee
where the zero mode Bogoliubov operator $G_0$ has the following form
\be
G_0 = - i \theta_0 (\b_T )\sum\limits_{\m = 1 }^{D-1}
\left( \tilde{\a}^{\m}_{0} \a^{\m}_{0} - \a^{\dagger\m}_{0}\tilde{\a}^{\dagger\m}_{0} \right),
\label{Bogoliubovzero}
\ee
and the parameter $\theta_0$ is related to the distribution function by the following relation
\be
\cosh \theta_{0}(\b_T ) = \left( 1 - e^{-\b_T \o_0} \right)^{-\f{1}{2}}.
\label{zeromodedistrib}
\ee
Here, the zero mode frequency is 
\be
\o_0 = \frac{\pi m {\a '}^2}{l}.
\label{zerofreq}
\ee
The operators $G$ and $\overline{G}$ are the Bogoliubov operators for $\a$ and $\b$ modes, respectively, and take the following form
\bea
G_{0} & = & \sum\limits_{n=1}^{\infty} G_n = -i\sum\limits_{n=1}^{\infty} \theta_n (\b_T )
\sum\limits_{\m = 1 }^{D-1}
\left( \tilde{\a}^{\m}_{n} \a^{\m}_{n} - \a^{\dagger\m}_{n}\tilde{\a}^{\dagger\m}_{n} \right),
\label{Gop}\\
\overline{G}_{0} & = & \sum\limits_{n=1}^{\infty} \overline{G}_n = -i\sum\limits_{n=1}^{\infty} \overline{\theta}_n (\b_T )
\sum\limits_{\m = 1 }^{D-1}
\left( \tilde{\b}^{\m}_{n} \b^{\m}_{n} - \b^{\dagger\m}_{n}\tilde{\b}^{\dagger\m}_{n} \right). 
\label{barGop}
\eea
The coefficients $\theta_n (\b_T ) = \overline{\theta}_n (\b_T )$ are equal for all $n = 1, 2, \ldots$ and $\m = 1, 2, \ldots , D-1$ since the oscillators are identical in both sectors and along all directions of the transverse tangent space. Their relation with the bosonic distribution is given by the following relation
\be
\cosh \theta_{n}(\b_T ) = \cosh \overline{\theta}_{n}(\b_T ) = \left( 1 - e^{-\b_T \o_n} \right)^{-\f{1}{2}},
\label{nmodedistrib}
\ee
where
\be
\o_n = \overline{\o}_n = \pi \a ' n \left( \frac{\O^{2}_{n} + 1}{\O_n} \right).
\label{nfreq}
\ee
The thermal vacuum obtained in (\ref{thermvacphs}) is the image of the total vacuum given by the following relation
\be
|0\rangle\rangle \equiv |0\rangle \tilde{|0\rangle} =
|0\rangle\rangle_0 \otimes_{n} |0\rangle\rangle_n \otimes_{n}\overline{|0\rangle\rangle}_n ,
\label{totalvac}
\ee
under the unitary transformation generated by the Bogoliubov operator
\be
|0(\b_T ) \rangle\rangle = e^{-i\GG}|0\rangle\rangle.
\label{thermvacmap}
\ee 
Since $|0\rangle\rangle $ belongs to the total physical space, the Bogoliubov operator maps $\hat{\HH}_{phys}$ to the thermal Hilbert space $\HH (\b_T )$. The total vacuum is annihilated by all annihilation operators and it is translational invariant. 

The thermal vacuum (\ref{thermvacmap}) can be defined in the same way if thermal operators are constructed by acting upon the set of all oscillator operators
\be
\OO \equiv \{ O \} =
\{ 
\a^{\m}_{0}, \a^{\dagger\m}_{0}, \tilde{\a}^{\m}_{0}, \tilde{\a}^{\dagger}_{0};
\a^{\m}_{n}, \a^{\dagger\m}_{n}, \tilde{\a}^{\m}_{n}, \tilde{\a}^{\dagger}_{n};
\b^{\m}_{n}, \b^{\dagger\m}_{n}, \tilde{\b}^{\m}_{n}, \tilde{\b}^{\dagger}_{n}
\},
\label{setosczeroT}
\ee
through the similarity transformations generated by the Bogoliubov operator
\be
\OO (\b_T ) = e^{-i\GG}\OO e^{i\GG} =  \{ e^{-i\GG} O e^{i\GG} \}.  
\label{setoscnonzeroT}
\ee
The space $\HH (\b_T )$ has a Fock space structure. The thermal vacuum state satisfies the relations
\bea
\a^{\m}_{0} (\b_T ) |0(\b_T ) \rangle\rangle & = & 
\a^{\m}_{n} (\b_T ) |0(\b_T ) \rangle\rangle =
\b^{\m}_{n} (\b_T ) |0(\b_T ) \rangle\rangle = 0,
\label{firstthermvacrel}\\
\tilde{\a}^{\m}_{0} (\b_T ) |0(\b_T ) \rangle\rangle & = & 
\tilde{\a}^{\m}_{n} (\b_T ) |0(\b_T ) \rangle\rangle =
\tilde{\b}^{\m}_{n} (\b_T ) |0(\b_T ) \rangle\rangle = 0.
\label{secondthermvacrel}
\eea

Since the Bogoliubov operator mixes zero temperature oscillating modes from all sectors, the finite temperature non-tilde and tilde oscillators do not represent anymore string and reservoir degrees of freedom, respectively. They rather describe thermal oscillations of the heated system which results from the interaction of zero temperature string and reservoir. A  thermal string state will contain an arbitrary number of thermal excitations from all sectors and it general form is given by the following relation
\bea
& & |\Psi^{\m_1 \ldots \n_1 \ldots \rho_1 \ldots \t_1}_{m_1 \ldots \n_1 \ldots p_1 \ldots q_1}(\b_T)\rangle\rangle =\nonumber\\
& & \left[ \a^{\dagger\m_1}_{m_1}(\b_T) \right]^{k^{\m_1}_{m_1}} \cdots
\left[ \b^{\dagger\n_1}_{n_1}(b_T) \right]^{\overline{k}^{\n_1}_{n_1}} \cdots
\left[ \tilde{\a}^{\dagger\rho_1}_{p_1}(\b_T) \right]^{s^{\rho_1}_{p_1}}
\cdots
\left[ \tilde{\b}^{\dagger\t_1}_{q_1}(\b_T) \right]^{\overline{s}^{\t_1}_{q_1}}
\cdots
|0(\b_T )\rangle\rangle.
\label{genthermstate}
\eea
The above state contains, $k^{\m_1}_{m_1}$ thermal excitations of type $\a_{m_1}$ in the direction $\m_1$, $\overline{k}^{\n_1}_{n_1}$ thermal excitations of type $\b_{n_1}$ in the direction $\n_1$ etc.

\subsection{Constraints of the Thermal String in $D=2+1$}

The symmetries of the thermal string should be checked by verifying the conformal algebra on the thermal states. However, it is a simple exercise to show that the operators $L_n$ and $\overline{L}_n$ do not commute with the Bogoliubov operator. Thus, the conformal algebra is broken at finite temperature. Nevertheless, it is natural to ask whether there are any symmetries and constraints left that should be imposed on the thermal string states obtained in the previous section. To answer this question, note that the string dynamics at finite temperature can be derived from the following Lagrangian
\be 
\LL_{2}(\b_T ) = e^{-i\GG}\hat{\LL}_{2}e^{i\GG},
\label{fintemplagr}
\ee
where $\hat{\LL}_2 = \LL_2 - \tilde{\LL}_2$, $\LL_2$ is the Lagrangian corresponding to the truncated action given in the relation (\ref{truncatedaction}) and $\tilde{\LL}_2$ is its reservoir counterpart. From the above Lagrangian one can build the following Hamiltonian and the world-sheet momentum  
\be
\hat{H} = H - \tilde{H}~~,~~\hat{P} = P - \tilde{P},
\label{tothamtotmom}
\ee
and show that the following commutation relations hold
\be
\left[ \hat{H}, \GG \right] = \left[ \hat{P}, \GG \right] = 0.
\label{commhammom}
\ee
Since $\hat{H}$ is the total Hamiltonian of the bosonic string, one can interpret $\hat{P}$ as being its total momentum. Thus, any physical state $| \Psi_{phys}\rangle\rangle =|\Psi_{phys}\rangle \widetilde{|\Psi_{phys}\rangle}$ can be mapped to the following thermal state
\be
|\Psi_{phys}(\b_T ) \rangle\rangle = e^{-i\GG} | \Psi_{phys}\rangle\rangle,   
\label{physstatesatT}
\ee
the latter being invariant under the world-sheet translation of the total string
\be
\hat{P}|\Psi_{phys}(\b_T ) \rangle\rangle = 0.
\label{invarphys}
\ee
The above relation is on equal footing of importance with the Hamiltonian invariance and therefore it can be used as the definition of the thermal physical states. Note that the operators in (\ref{invarphys}) are at zero temperature while the state are at finite temperature. It is a simple exercise to write (\ref{invarphys}) in terms of objects at finite temperature only.

\section{Thermodynamical Functions of the Thermal String}

In this section we are going to compute the entropy and the free energy of the bosonic string at finite temperature. The basic hypothesis is that the semiclassical thermal string is at equilibrium in the free falling reference system of the center of mass. Then the free thermal string excitations are created out of the thermal equilibrium vacuum $|0(\b_T ) \rangle \rangle$ by the temperature dependent creation operators and the average string energy $\langle H \rangle$ is equal to the $\langle \tilde{H} \rangle$ \cite{tu,ubook}. Thus the thermal equilibrium states can be understood as states in which pairs of tilde and non-tilde excitations fluctuate such that when an excitation of one type is created another excitation of the other type is annihilated and subjected to the relation (\ref{invarphys}) from the previous section. 

Consider the string oscillation modes in the physical Hilbert space as discussed in the previous section. Then the string entropy in $k_B$ units can locally be defined as the expectation value of the following operator in the thermal vacuum \cite{ubook}
\bea
K &=& - \sum\limits_{n=1}^{\infty}\sum\limits_{\m=1}^{D-1}
\left[ 
	\left( 
		\a^{\mu \dagger}_{n} \a^{\m}_{n} + \b^{\m\dagger}_{n} \b^{\m}_{n}
	\right)
				\log\sinh^2\theta_{n}(\b_T) -
	\left( 
		\a^{\m}_{n} \a^{\mu \dagger}_{n} + \b^{\m}_{n} \b^{\m\dagger}_{n} 
	\right)
				\log\cosh^2\theta_{n}(\b_T)			
\right]\nonumber\\
&-&
\sum\limits_{\m=1}^{D-1}
\left[ 
		\a^{\mu \dagger}_{0} \a^{\m}_{0} \log\sinh^2\theta_{0}(\b_T) -
		\a^{\m}_{0} \a^{\mu \dagger}_{0} \log\cosh^2\theta_{0}(\b_T)			
\right].
\label{entropyop}
\eea 
By using the expectation value of the zero temperature number operator in the thermal vacumm 
\bea
\langle\langle 0 (\b_T ) | \a^{\mu \dagger}_{n} \a^{\m}_{n} |0 (\b_T ) \rangle\rangle =
\langle\langle 0 (\b_T ) | \b^{\mu \dagger}_{n} \b^{\m}_{n} |0 (\b_T ) \rangle\rangle =
\sinh^2 \theta_{n}(\b_T ),
\label{numbopexp}
\eea
for all oscillators, we can write the entropy of the bosonic string in the following form
\bea
S &=& 2(D-1)k_B \sum\limits_{n=1}^{\infty}
\left[
			\b_T \pi \a ' n \o_n f(\pi \a ' n \o_n ) + \log(1+f(\pi \a ' n \o_n))
\right]
\nonumber\\
&+&
(D-1)k_B \left[
							\b_T \frac{\pi m {\a '}^2}{l}f(\frac{\pi m {\a '}^2}{l}) + \log \left(1+
							 f(\frac{\pi m {\a '}^2}{l}) \right)
				 \right],	
\label{entropy}
\eea
where
\be
f(\o_n) = \frac{1}{e^{\b_T \o_n}-1},
\label{thetadistrib}
\ee
is the bosonic distribution function for all $n= 0,1,2,\ldots$. Now let us compute the thermal string free energy. By definition, its general expression is given by the expected value in the thermal vacuum of the following operator 
\be
F = -\frac{1}{k_B}K + H.
\label{freeenergy}
\ee
By using the above calculations and the explicit form of the local Hamiltonian we arrive at the following form of the free energy
\bea
\langle \langle F \rangle  \rangle &=& (D-1)\sum\limits_{n=1}^{\infty}
				 \left[ 	
				 4\pi \a ' n\o_n f(\pi \a ' n \o_n) + \frac{2\pi m {\a '}^2}{l}
				 f(\frac{\pi m {\a '}^2}{l})
				 \right]
\nonumber\\
&+&
\frac{(D-1)}{\b_T} \left[
				 2 \sum\limits_{n=1}^{\infty}
				 	\log\left(1 + f(\pi \a ' n \o_n)\right) +
				 	\log\left(1 + f(\frac{\pi m {\a '}^2}{l}) \right)
				 	\right]
				 	- \pi m^2 {\a '}^2
				 ,				 
\label{freeenergycalc}
\eea
where the last two terms represent the contribution of the zero modes and string mass, respectively. 

The factor of 2 in the formula (\ref{entropy}) denotes the contribution of $\a$ and $\b$ oscillators for $n>1$, respectively. The last two terms in the entropy define a function $S_0$ on the string tension and the cosmological constant. At constant temperature and 
\be
T^{2}_{s} >> \f{m \b_T }{4 \pi} \sqrt{-\L},
\label{inftylimittension}
\ee 
$S_0$ depends on temperature as
\be
S_0 \approx 1  + \log\left( 1+ \frac{1}{\b_T \o_0} \right)
\label{approx1}
\ee
Thus, we can expect this contribution be relevant at high temperatures
\be
T >> \frac{m}{4 \pi k_B}\sqrt{-\L}. 
\label{hightemp}
\ee
At values of the string constant  
\be
T^{2}_{s} << \f{m \b_T }{4 \pi} \sqrt{-\L},
\label{zerolimittension}
\ee 
the last two terms in the entropy depend on the temperature as
\be
S_0 \approx \log(2 - \b_T \o_0) + \b_T \o_0 -(\b_T \o_0 )^2.
\label{approx2}
\ee
In this case there is a critical temperature 
\be
T_c = \frac{8\pi k_B}{m T^{2}_{s}\sqrt{-\L}}.
\label{criticaltemp}
\ee
For $T < T_c$ the zero mode entropy is no longer well defined.
This result may be interpreted as a failure of the semiclassical quantization procedure in the tensionless limit of the string theory. In this limit, the interaction among string oscillators may induced effects beyond the first order approximation of the perturbative expansion in $\e$. The string effects are reduced in the large tension limit in which the string behaves more like a particle. Similar conclusions can  be drawn for the string free energy.

\section{Conclusions and Discussions}

In this paper we have constructed the locally free thermal semiclassical string excitations in the AdS spacetime in the presence of the exact metric. The semiclassical quantization allows to study the free oscillation modes which appear at the the first order in the power expansion of the string fields in the dimensionless parameter $\e$. From physical point of view, this expansion corresponds in $D=2+1$ black-hole AdS spacetime to the expansion in $\o /M$ where $\o$ is a typical string oscillation frequency scale and $M$ is the black-hole mass. Therefore, the approximation holds for strings in strong gravitational field \cite{ns2}. The truncated theory at zero temperature displays the conformal invariance and has the level matching condition constraint which define the physical subspace of the Hilbert space. By fixing the gauge symmetries of the theory we were able to map the string theory from $T = 0$ to $T \neq 0$ 
by a Bogoliubov operator constructed within the TFD formalism. To this end, we have assumed that the semiclassical string is in contact with a thermal reservoir in the chosen reference frame. Then application of the TFD to the physical Hilbert space of the local transverse oscillation to the geodesic is quite straightforward. Also, in this setup we have calculated the entropy and the free energy of the thermal string.  

The relations obtained in this paper are valid for the free thermal string in arbitrary AdS spacetime just in the first order approximation of the semiclassical quantization. In that case
there is no conformal symmetry of the vacuum. However, as pointed out in \cite{ns6}, the level matching condition and the Hamiltonian should be the same in this approximation. For arbitrary dimension of the AdS, our result show that there are terms in the entropy that depend on the cosmological constant $\L$. Also, there is a critical temperature under which the thermalization based on the semiclassical quantization is not well defined in the large string tension limit, i. e. away from the particle approximation of the string theory.

Our results were derived in the physical Hilbert space. However, it is possible to work in the full Hilbert space, too, in which case the level matching condition will show up in the corresponding formulas explicitly. The thermal partition function $Z(\b_T)$ has the following form in the full Hilbert space  
\be
Z(\b_T ) = \frac{\exp (\b_T \pi m^2 {\a '}^2)}
			{
				\left[ 
							1 - \exp
											\left( 
													-\b_T \frac{\pi m {\a '}^2}{l}
											\right)
				\right]^{D-1}
			}
\int\limits_{-1/2}^{+1/2} ds \prod\limits_{n=1}^{\infty}
				\left[
					\left(
					1-e^{\pi \a ' n \l_n(\b_T,s)}
					\right)
				\left(
					1-e^{\pi \a ' n \overline{\l}_n(\b_T,s)}
					\right)
				\right]^{1-D}.
\label{fullHpartfunct}
\ee
The thermal vacuum state has the following expansion in terms of the full Hilbert space basis 
\bea
|0(\b_T ) \rangle\rangle & = & 
\frac{ Z^{-\frac{1}{2}}(\b_T)
\exp (\frac{\b_T \pi m^2 {\a '}^2)}{2}}
			{
				\left[ 
							1 - \exp
											\left( 
													-\b_T \frac{\pi m {\a '}^2}{l}
											\right)
				\right]^{\frac{D-1}{2}}
			} 
		\sum\limits_{w}
		\sum\limits_{\overline{w}}
	\left[
	\int\limits_{-1/2}^{+1/2} ds
		\exp 
			\left(
				i\pi n s \sum\limits_{\m=1}^{D-1} 
					\left(
					k^{\m}_{n} - \overline{k}^{\m}_{n}
					\right)
			\right)		
	\right]^{-\frac{1}{2}}
\nonumber\\
&\times &		
\exp 
		\left(
		-\b_T \pi \a ' \sum\limits_{n=1}^{\infty}n\o_n
									 \sum\limits_{\m=1}^{D-1}
									 \left(
									 k^{\m}_{n} + \overline{k}^{\m}_{n}
									 \right)	
		\right)	
|w,\overline{w}\rangle	\widetilde{|w,\overline{w}\rangle},	
\label{fullthermvacuum}
\eea
where the partition function is now the one given in (\ref{fullHpartfunct}). To compute the entropy and the free energy  one has to take the expectation value of the entropy operator $K$ and the Hamiltonian in the state (\ref{fullthermvacuum}) and to perform the integral over $s$ parameter, too. Note that in the full Hilbert space the Hamiltonian has the form  
\be
H' =  2\pi\a ' \sum\limits_{n \geq 1} 
\left[
	\left( 
				\frac{\O^{2}_{n} + 1}{\O_n} 
	\right)  
	\left( N_n + \overline{N}_{n} \right)
	+
	2\pi i s 
	\left( 
				N_n - \overline{N}_n 
	\right)
\right]
+\frac{\pi m {\a '}^2}{l} \a^{\dagger}_{0} \cdot \a_{0} - 
\pi m^2{\a '}^2 ,
\label{fullHam}
\ee
which differs from (\ref{hamiltonian}) in that the level matching constraint has been added to it. This is in agreement to the interpretation of the parameter $s$ as a Langrange multiplier \cite{ng3}. However, working in the full Hilbert space implies manipulating the Columbeau's generalized functionals \cite{c}. It is an interesting problem to show that the two formulations coincide.

{\bf Acknowledgments}
H. B. and M. A. S. would like to thank to M. T. Orlando for discussions. I. V. V. would like to thank to J. A. Helay\"{e}l-Neto, S. A. Dias and A. M. O. de Almeida for discussions and  hospitality at LAFEX-CBPF where part of this work was being carried out.


\begin{thebibliography}{99}
\bibitem{brfw}
  J.~Balog, L.~O'Raifeartaigh, P.~Forgacs and A.~Wipf,
  Nucl.\ Phys.\ B {\bf 325}, 225 (1989).
\bibitem{dpl}
  L.~J.~Dixon, M.~E.~Peskin and J.~D.~Lykken,
  Nucl.\ Phys.\ B {\bf 325}, 329 (1989).
\bibitem{jm}
  J.~M.~Maldacena,
  Adv.\ Theor.\ Math.\ Phys.\  {\bf 2}, 231 (1998).
\bibitem{ns1}
  H.~J.~de Vega and N.~Sanchez,
  Phys.\ Lett.\ B {\bf 197}, 320 (1987).
\bibitem{ns2}
  H.~J.~de Vega and N.~Sanchez,
  Nucl.\ Phys.\ B {\bf 299}, 818 (1988).
\bibitem{nsa}
  N.~G.~Sanchez,
  Phys.\ Lett.\ B {\bf 195}, 160 (1987).
\bibitem{ns3}
  A.~L.~Larsen and N.~Sanchez,
  Phys.\ Rev.\ D {\bf 50}, 7493 (1994).
\bibitem{ns4}
  H.~J.~de Vega, A.~L.~Larsen and N.~G.~Sanchez,
  Phys.\ Rev.\ D {\bf 58}, 026001 (1998)
\bibitem{bhtz}
  M.~Banados, M.~Henneaux, C.~Teitelboim and J.~Zanelli,
  Phys.\ Rev.\ D {\bf 48}, 1506 (1993).
\bibitem{tu}
  Y.~Takahasi and H.~Umezawa,
  Collect.\ Phenom.\  {\bf 2}, 55 (1975).
\bibitem{ubook}H. Umezawa, {\em Advanced Field Theory: Micro, Macro and Thermal Physics}, (AIP New-York, 1993). 
\bibitem{ns6}
  A.~L.~Larsen and N.~Sanchez,
  Phys.\ Rev.\ D {\bf 52}, 1051 (1995).
\bibitem{btz}
  M.~Banados, C.~Teitelboim and J.~Zanelli,
  Phys.\ Rev.\ Lett.\  {\bf 69}, 1849 (1992)
\bibitem{yl1}
  Y.~Leblanc,
  Phys.\ Rev.\ D {\bf 36}, 1780 (1987).
\bibitem{yl2}
  Y.~Leblanc,
  Phys.\ Rev.\ D {\bf 37} 1547 (1988).
\bibitem{yl3}
  Y.~Leblanc,
  Phys.\ Rev.\ D {\bf 38}, 3087 (1988).
\bibitem{yl4}
  Y.~Leblanc,
  Phys.\ Rev.\ D {\bf 39} 1139 (1989).
\bibitem{yl5}
  Y.~Leblanc, M.~Knecht and J.~C.~Wallet,
  Phys.\ Lett.\ B {\bf 237} 357 (1990).
\bibitem{yl6}
  Y.~Leblanc,
  Phys.\ Rev.\ Lett.\  {\bf 64} 831 (1990).
\bibitem{fns1}
  H.~Fujisaki, K.~Nakagawa and I.~Shirai,
  Prog.\ Theor.\ Phys.\  {\bf 81}, 570 (1989).
\bibitem{fn1}
  H.~Fujisaki and K.~Nakagawa,
  Prog.\ Theor.\ Phys.\  {\bf 82}, 236 (1989).
\bibitem{fn2}
  H.~Fujisaki and K.~Nakagawa,
  Prog.\ Theor.\ Phys.\  {\bf 82}, 1017 (1989).
\bibitem{fn3}
  H.~Fujisaki and K.~Nakagawa,
  Europhys.\ Lett.\  {\bf 14}, 737 (1991).
\bibitem{fn4}
  H.~Fujisaki and K.~Nakagawa,
  Europhys.\ Lett.\  {\bf 20}, 677 (1992).
\bibitem{fn5}
  H.~Fujisaki,
  Europhys.\ Lett.\  {\bf 28}, 623 (1994).
\bibitem{f1}
  H.~Fujisaki,
  Europhys.\ Lett.\  {\bf 39}, 479 (1997).
\bibitem{ivv7}
  M.~C.~B.~Abdalla, E.~L.~Graca and I.~V.~Vancea,
  Phys.\ Lett.\ B {\bf 536}, 114 (2002).
\bibitem{ng1}
  D.~L.~Nedel, M.~C.~B.~Abdalla and A.~L.~Gadelha,
  Phys.\ Lett.\ B {\bf 598}, 121 (2004).
\bibitem{ng2}
  M.~C.~B.~Abdalla, A.~L.~Gadelha and D.~L.~Nedel,
  JHEP {\bf 0510}, 063 (2005).
\bibitem{ng3}
  M.~C.~B.~Abdalla, A.~L.~Gadelha and D.~L.~Nedel,
  Phys.\ Lett.\ B {\bf 613}, 213 (2005).
\bibitem{ng4}
  M.~C.~B.~Abdalla, A.~L.~Gadelha and D.~L.~Nedel,
  PoS {\bf WC2004}, 020 (2004).
\bibitem{ng5}
  M.~C.~B.~Abdalla, A.~L.~Gadelha and D.~L.~Nedel,
  PoS {\bf WC2004}, 032 (2004).
\bibitem{ivv1}
  I.~V.~Vancea,
  Phys.\ Lett.\ B {\bf 487}, 175 (2000).
\bibitem{ivv2}
  M.~C.~B.~Abdalla, A.~L.~Gadelha and I.~V.~Vancea,
  Phys.\ Rev.\ D {\bf 64}, 086005 (2001).
\bibitem{ivv3}
  M.~C.~B.~Abdalla, A.~L.~Gadelha and I.~V.~Vancea,
  Phys.\ Rev.\ D {\bf 66}, 065005 (2002).
\bibitem{ivv4}
  M.~C.~B.~Abdalla, A.~L.~Gadelha and I.~V.~Vancea,
  ``$D$-branes at finite temperature in TFD,''
  [arxiv:hep-th/0308114].
\bibitem{ivv5}
  M.~C.~B.~Abdalla, A.~L.~Gadelha and I.~V.~Vancea,
  Int.\ J.\ Mod.\ Phys.\ A {\bf 18}, 2109 (2003).
\bibitem{ivv6}
  M.~C.~B.~Abdalla, A.~L.~Gadelha and I.~V.~Vancea,
  Nucl.\ Phys.\ Proc.\ Suppl.\  {\bf 127}, 92 (2004).
\bibitem{ivv8}
  I.~V.~Vancea,
  Phys.\ Rev.\ D {\bf 74}, 086002 (2006).
\bibitem{ivv9}
  I.~V.~Vancea,
  PoS {\bf WC2006}, 036 (2006),
  [arxiv:hep-th/0609188].
\bibitem{fg}
  J.~J.~Friess and S.~S.~Gubser,
  Nucl.\ Phys.\ B {\bf 750}, 111 (2006).
\bibitem{kl}
  M.~Kruczenski and A.~Lawrence,
  JHEP {\bf 0607}, 031 (2006).
\bibitem{egp}
  J.~M.~Evans, M.~R.~Gaberdiel and M.~J.~Perry,
  Nucl.\ Phys.\ B {\bf 535}, 152 (1998).
\bibitem{c}J. F. Columbeau, {\em Elementary Introduction to New Generalized Functions},
North Holland, 1985.
\end{thebibliography}
\end{document}